\documentclass[sn-nature]{sn-jnl}

\usepackage{graphicx}%
\usepackage{multirow}%
\usepackage{amsmath,amssymb,amsfonts}%
\usepackage{amsthm}%
\usepackage{mathrsfs}%
\usepackage[title]{appendix}%
\usepackage{xcolor}%
\usepackage{textcomp}%
\usepackage{manyfoot}%
\usepackage{booktabs}%
\usepackage{algorithm}%
\usepackage{algorithmicx}%
\usepackage{algpseudocode}%
\usepackage{listings}%
\usepackage{natbib}
\usepackage{amssymb}
\usepackage{todonotes}
\usepackage{siunitx}
\usepackage{svg}
\usepackage{caption}
\renewcommand{\figurename}{Figure}
\usepackage{array} 
\newcolumntype{P}[1]{>{\centering\arraybackslash}p{#1}}
\newcolumntype{M}[1]{>{\centering\arraybackslash}m{#1}}

\NewDocumentCommand{\rot}{O{45} O{1em} m}{\makebox[#2][l]{\rotatebox{#1}{#3}}}%
\usepackage{ltablex}
\usepackage{graphicx}
\usepackage{cleveref}
\usepackage{caption}
\usepackage{enumitem,kantlipsum}
\usepackage{multirow}
\usepackage{siunitx}
\usepackage{rotating} 
\usepackage{color,soul}
\usepackage{float}
\usepackage[boxruled,algo2e]{algorithm2e}
\usepackage{algpseudocode}
\usepackage{subcaption}
  
\theoremstyle{thmstyleone}%
\theoremstyle{thmstyletwo}%

\theoremstyle{thmstylethree}%

\begin{document}

\title[Engines]{Evaluating Search Engines and Large Language Models for Answering Health Questions}

\author*[1]{\fnm{Marcos} \sur{Fernández-Pichel}}\email{marcosfernandez.pichel@usc.es}

\author[1]{\fnm{Juan C.} \sur{Pichel}}\email{juancarlos.pichel@usc.es}

\author[1]{\fnm{David E.} \sur{Losada}}\email{david.losada@usc.es}

\affil[1]{\orgdiv{Centro Singular de Investigación en Tecnoloxías Intelixentes (CiTIUS)}, \orgname{Universidade de Santiago de Compostela}, \orgaddress{\city{Santiago de Compostela}, \postcode{15782}, \state{Galicia}, \country{Spain}}}

\abstract{Search engines (SEs) have traditionally been primary tools for information seeking, but the new Large Language Models (LLMs) are emerging as powerful alternatives, particularly for question-answering tasks. This study compares the performance of four popular SEs, seven LLMs, and retrieval-augmented (RAG) variants in answering 150 health-related questions from the TREC Health Misinformation (HM) Track. Results reveal SEs correctly answer 50–70\% of questions, often hindered by many retrieval results not responding to the health question. LLMs deliver higher accuracy, correctly answering about 80\% of questions, though their performance is sensitive to input prompts. RAG methods significantly enhance smaller LLMs’ effectiveness, improving accuracy by up to 30\% by integrating retrieval evidence.}


\maketitle

\section*{Introduction}

Recent progress in Natural Language Processing (NLP)
has positioned 
Large Language Models 
as major players in numerous 
Information Access tasks ~\cite{longpre2023flan,bubeck2023sparks,touvron2023llama}. The release of ChatGPT in November 2022 has been a game-changer globally, marking a significant milestone and revolutionizing many sectors. One of the outstanding features of current LLMs is their ability to generate coherent and human-like text,
which has garnered attention and excitement among practitioners, researchers, and the general public. 
This breakthrough has precipitated a transformative shift in the orientation of information access research towards LLMs, their potential applications, and the interconnection
between LLMs
and other computer-based tools. The conversational paradigm has gained traction, enabling more interactive and user-friendly search experiences \cite{mao2023large,friedman2023leveraging,polak2023extracting,o2022massive}; and many citizens currently turn to conversational AIs based on LLMs for consulting multiple types of information needs. 
However, traditional search 
still plays a crucial role in the generative AI era \cite{hersh2024search}. 
While LLMs may support advanced information access and reasoning capabilities, there is still a need to advance retrieval technologies. 
The role of traditional web search engines in answering
user-submitted queries is far from being relegated. 
Web search is widely used to obtain health advice~\cite{wu2024investigating}
and effective medical information retrieval has attracted the attention of the scientific community over the years~\cite{sivarajkumar2024clinical}. 
For example, Wang et al. evaluated several search engines in terms of their usability and effectiveness for searching for breast cancer information \cite{wang2012using}. They found that 
the results highly overlapped among the four search engines tested, all of them providing
rich information about breast cancer.
Zuccon and colleagues \cite{zuccon2015}
studied the effectiveness of search engines for the so-called ``diagnostic medical circumlocutory queries'', which are searches issued by individuals seeking information about their health using casual descriptions of symptoms rather than medical words.

The emergence and global adoption of advanced LLMs have sparked the urgent need to explore and understand 
their capacities and knowledge acquisition attributes. Some research studies have focused on the capabilities of these models under specific language understanding and reasoning benchmarks ~\cite{jiang2020can, liang2022holistic, chang2024survey}
and, specifically, 
interest in assessing 
the correctness of health-related AI-based completions has escalated. 
For example, Chervenak et al. demonstrated ChatGPT's abilities to answer fertility questions \cite{chervenak2023promise} 
and Duong and Solomon \cite{duong2023analysis} 
evaluated the effectiveness 
of LLMs
in comparison to humans 
when tasked with answering multiple-choice questions on human genetics. Similarly, 
Holmes et al. \cite{holmes2023evaluating} conducted a comparative study of 
LLMs' knowledge on the highly specialized subject of radiation oncology physics. Recently, Elgedawy and colleagues tested the capacity of LLMs to query an extensive volume of clinical records ~\cite{elgedawy2024dynamic} and 
Kim et al. assessed ChatGPT's accuracy in answering 57 epilepsy-related questions \cite{kim2024assessing}. Kim and others ~\cite{kim2024large} examined the diagnostic accuracy of GPT-4 compared to mental health professionals and
other clinicians employing clinical vignettes of obsessive-compulsive disorder (OCD).
In other papers, the authors reported studies on the role of LLMs for biomedical tasks~\cite{jahan2023evaluation}, patient-specific EHR questions~\cite{hamidi2023evaluation}, and bariatric surgery topics~\cite{samaan2023assessing}. Tang et al. \cite{tang2023evaluating} evaluated LLMs' ability to perform zero-shot medical summarization across six clinical domains.  With a broader perspective, 
other authors involved physicians in a thorough evaluation of the accuracy of ChatGPT in answering health queries \cite{johnson2023assessing}
or evaluated ChatGPT using the 
Applied Knowledge Test (AKT) from the Royal College of General Practitioners \cite{thirunavukarasu2023trialling}. 
All of the aforementioned studies are restricted to a single model, usually ChatGPT, and/or to a specific medical area. 
Kusa et al. \cite{kusa-etal-2023-dr} analyzed the impact of user's beliefs and prompt formulations on completions related to health diagnoses. The study explored the sensitivity of two GPT models to variations in user's context. Caramancion \cite{caramancion2024large} explored users' preferences between
search engines and LLMs. This analysis revealed interesting trends in user preferences, indicating a distinct tendency for participants to choose search engines for straightforward, fact-based questions. In contrast, LLMs were more frequently favored for tasks needing detailed comprehension and language processing.
The results clearly showed that 
 users prefer to search for medical information using traditional search engines. 
 Oeding and colleagues \cite{oeding2024chatgpt} compared GPT-4's and Google's results for searches concerning the Latarjet procedure (anterior shoulder instability). These authors discovered that GPT-4 provided more information based on academic sources than Google in response to the patient's queries.

In the area of retrieval-augmented models for the health domain, Li et al. \cite{li2023chatdoctor} fine-tuned Llama with medical conversations
and injected medical evidence from Wikipedia and other medical sources, while 
Koopman and Zuccon \cite{koopman2023dr} evaluated ChatGPT's capacity to answer health questions (based solely on its internal knowledge or, alternatively, fed with offline retrieval evidence). Xiong et al. \cite{xiong2024benchmarking} proposed MedRAG, a system that indexes multiple medical corpora and retrieves relevant evidence to ground different LLMs. This study found that simpler models can reach the performance of GPT-4 when grounded with relevant medical information.

In line with these developments, it is 
crucial to acknowledge the significance of accurate health information
and there is a pressing need to evaluate 
the abilities of classic and new information access tools 
in answering medical questions. There is a lack of comprehensive studies that compare the effectiveness of LLMs with that of traditional SEs in the context of health information seeking.
Furthermore, given that the input prompt significantly influences the efficacy of LLMs
\cite{jiang2020can,brown2020language,liu2023pre}, understanding models' effectiveness 
with different types of prompts 
is of utmost importance. 
It is also crucial to explore the effect of combining both classes of tools and, for example, 
explore the behavior of LLMs
when prompted with medical questions together with related search results. 
This paper aims to contribute towards filling these gaps by conducting a thorough study aimed 
at responding to the following research questions:

\begin{itemize}
    \item RQ1. To what extent do search engines retrieve results that help to answer medical questions?  Does the correctness of the information provided drop as 
    we go down the search engine result page?
    \item RQ2. Are LLMs reliable in providing accurate medical answers? How do different models compare regarding their effectiveness in responding to medical questions?
    \item RQ3. Does the given context influence the capabilities of LLMs in providing correct answers? 
    Is there an observable improvement in these models when they are exposed to a few in-context examples?
    \item RQ4. Do LLMs improve their performance when fed with web retrieval results? 
\end{itemize}

Our comparison of SEs included
experiments with Google, Bing, Yahoo! and
DuckDuckGo. We found that Bing seems to be the most solid choice. However, it is not significantly better than the others. 
Our evaluation also suggests that extracting the answers from the top ranked webpage often produces
good results. This is good news, as web users are known to be reluctant to inspect many items from the search engine result pages (SERPs). 
Moreover, LLMs show an overall good performance, but they are still very sensitive to the input prompt, and, in some instances, they provide highly inaccurate responses. Finally, augmenting LLMs with 
retrieval results from SEs looks promising and, according to our experiments, even the smallest 
LLMs can achieve top-tier performance 
when provided with suitable retrieval evidence.
These findings significantly contribute to our understanding of health information seeking using the new LLMs and traditional search engines.

\section*{Results}

In this section, we report the effectiveness of different search engines in providing correct responses to binary health questions and analyze web user experience following two user models (see Methods). 
We also evaluate the performance of LLMs when prompted with the same health questions (under zero- and few-shot settings), propose a taxonomy of errors made by LLMs, and present results for retrieval-augmented strategies. 

\begin{figure}[t!]
  \centering
  
  \includegraphics[width=\textwidth]{figures/Figure_1.png}
\caption[Figure]{}
  
  \label{fig:cum-accuracy-all}
\end{figure}

\subsection*{Search Engines} \label{sec:se-res}

Figure~\ref{fig:cum-accuracy-all} shows the results for three different collections of questions. 
From these plots, we observe no significant drop in performance as we move down the rankings, which speaks well of the SEs' retrieval capabilities. With respect to different engines, Bing seems a solid choice in all datasets. However, we ran pairwise comparisons between SEs
(Mann-Whitney U test, $\alpha=0.05$), for the cutoff positions 1 and 20
(i.e., statistical differences
between their P@1 or their P@20 values), and found no significant differences
between any pair of engines. 
This suggests that the SEs
have similar performance and,
according to our experiments,
we cannot declare a clear
winner. 
By comparing the plots for the three collections, we can observe that the 2022 topics produced the best results. A possible explanation for this outcome is that the 2022 collection contains fewer specialized health questions, and it might be easier to retrieve correct answers from the web. 

\begin{figure}[t!]
    \centering
    \makebox[\linewidth]{\includegraphics[width=0.6\linewidth]{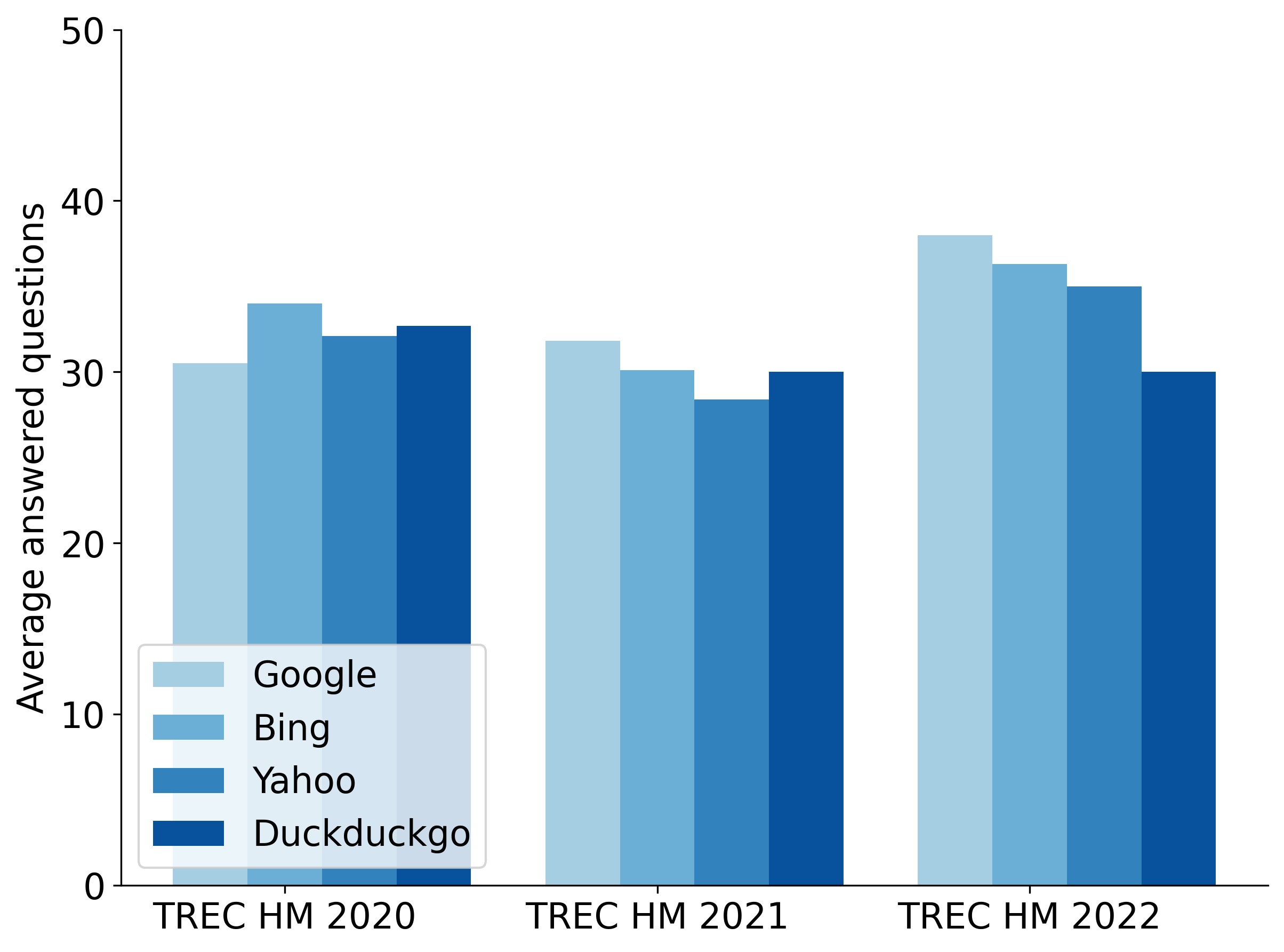}} 
    \caption{}
    \label{fig:ans}
\end{figure}

We also wanted to analyze the extent to which
retrieval results provide an actual
answer to the health questions. Ranked lists contain off-topic results and, additionally, many on-topic pages do not provide a clear response to the user's request. For each of the 50 health questions available in each collection, we processed 
its top retrieved results to
identify whether they answered 
the health question. This identification was automatic, supported by passage extraction and reading comprehension technology (see Methods). 
For example, for Google's rankings and n=1 we computed the number of entries that actually provided a response to the health query. The same count was obtained for each position (up to rank \#20) and, finally, a global answering score was obtained by averaging the number of questions that were actually answered at each position.
These scores, plotted in Figure~\ref{fig:ans}, are much lower than 
the number of questions in these collections (50),
reflecting that many retrieval results
do not provide a clear yes/no recommendation to the health question. 
According to this experiment, 
Bing is the engine whose rankings provided more answers 
for 2020 questions (around 30 questions answered) and, from Figure \ref{fig:cum-accuracy-all}(a) we infer that Bing's responses tend to be slightly better than those
provided by the three other SEs. 
For 2021, Google and DuckDuckgo supplied responses to more questions but at the cost of providing more incorrect responses throughout the ranking (Figure \ref{fig:cum-accuracy-all}(b)). Finally, for 2022 questions, Google provided many responses
and, according to Figure \ref{fig:cum-accuracy-all}(c), it was not inferior to the other SEs. 
Note also that, on average, 
between 15 and 20 questions 
did not receive a response at any given ranked position. This suggests
that restricting the analysis to a few top ranked webpages, which is the common behavior of end users, often 
results in a failure to meet these health
information needs.

The  precision of the SEs is low (60\%-70\%), but the good news 
is that this is because many ranked results
did not provide an answer. 
In fact, if we do not count the entries that do not provide a response as failures, then precision rises to 
80\%-90\% for all search engines. Although high, these values still reflect a worrying percentage of 10-15\% of incorrect responses in the search engine results pages.

\begin{figure}[t!]
 
    \centering
    \includegraphics[width=0.6\textwidth]{figures/Figure_3.png}
  
\caption{}
\label{fig:user-heuristic}
\end{figure}

Let us now evaluate the search results using two user behavior models: i) lazy, modeling a user who stops inspecting results when presented with the first entry that gives a yes/no response to his question, and ii) diligent, modeling a user who only stops after contrasting three responses (taking a majority vote decision, see Methods). 
In Figure~\ref{fig:user-heuristic}, we report the percentage of correct decisions, incorrect decisions, and no decisions for the three collections. Note that in a few instances, the user exhausted the ranked list and found no response (grey area in Figure~\ref{fig:user-heuristic}). These cases are more frequent in the diligent user scenario because this model needs three answers to make an informed decision. We also report in Table~\ref{tab:effort}
the effort required to make a decision, measured as the mean number of results a user needs to inspect before reaching the stopping criterion. 

Observe that the lazy behavior produces similar or even better results, requiring less effort, 
compared with those achieved by diligent behavior. 
Indeed, the additional effort spent by the diligent user
does not translate into a lower percentage of incorrect decisions. 
This may sound surprising, but it 
somehow reinforces the 
confidence in the search engine top 
result (the top answer suffices). 
In fact, our results suggest
that the diligent user, who
goes deeper in the ranking to find additional responses, would make poorer health-related decisions.

\begin{table}[t!]
    \centering
    \begin{subtable}{\textwidth}
        \centering
        \begin{tabular}{l|cc|cc|cc|}
            & \multicolumn{2}{c|}{\textbf{TREC HM 2020}} & \multicolumn{2}{c|}{\textbf{TREC HM 2021}} & \multicolumn{2}{c|}{\textbf{TREC HM 2022}} \\
            & \emph{Lazy} & \emph{Diligent} & \emph{Lazy} & \emph{Diligent} & \emph{Lazy} & \emph{Diligent}  \\ 
            \hline\hline
            Google     & 1.7 & 5.2 & 1.7 & 5.2 & 2 & 4.9 \\
            Bing       & 1.8 & 5.4 & 1.5 & 4.3 & 1.6 & 4.6 \\
            Yahoo      & 1.6 & 5.5 & 1.6 & 5.2 & 1.7 & 4.4 \\
            Duckduckgo & 1.7 & 5.1 & 1.4 & 4.9 & 1.7 & 4.6 \\
            \hline \hline
        \end{tabular}
    \end{subtable}
  \caption{Effort required to make a decision by each user behavior model. The effort is measured as the mean number of results inspected.}
    \label{tab:effort}
\end{table}

\subsection*{Large Language Models} \label{sec:llm-res}

Next, we evaluate the performance of LLMs for our binary question-answering task under zero- and few-shot settings. We also propose a taxonomy for the errors that these models tend to produce and perform an error analysis based on this taxonomy. Finally, the results for retrieval-augmented strategies are presented.

\begin{figure}[t!]

    \centering
    \includegraphics[width=\textwidth]{figures/Figure_4.png}
  
  \caption{}
  \label{fig:zeroshot}
\end{figure}

Figure~\ref{fig:zeroshot} plots the 0-shot accuracy for seven LLMs and three prompting strategies (``no context'', ``non-expert'' and ``expert'', see Methods). All models show remarkable performance for the TREC HM 2020 collection, with the exception of FlanT5.
In this collection, Llama3, MedLlama3 and text-davinci-003 stand out from the rest of LLMs. For the 2021 questions, the accuracy generally drops, but the same models dominate, with special mention to MedLlama3. Regarding the TREC HM 2022 data, ChatGPT and GPT-4  stand out. Additionally, there are variances in performance linked to the use of different prompts. In general, the ``expert'' context is the most effective, and it seems to be guiding the model toward 
more reputed knowledge sources.

The models generally show relative stability, but their performance still varies based on the inputs provided. For example, FlanT5 and 
text-davinci-002 are highly sensitive
to the type of prompt. 
This raises concerns, as a model's accuracy can drop from 90\%  to 75\% or even lower numbers. While the overall performance levels are high, these inconsistencies are troubling. Even when using the most reliable prompt (``expert''), there are still concerning situations. For instance, GPT-4's accuracy drops to 68\% on the 2021 dataset.

The difficulty levels of the three datasets vary. The 2020 health questions (centered on COVID-19) seem easier for the large language models. 
A possible reason could be the significant relevance of COVID-19 as a topic, potentially encouraging specialized data curation processes.

We ran McNemar's test to ascertain the statistical significance of the pairwise performance disparities between the leading models. The comparison between GPT-4 and ChatGPT showed no significant difference in 7 out of 9 cases (3 datasets x 3 prompts). The comparison between Llama3 --or its fine-tuned version, MedLlama3-- and ChatGPT revealed no significant difference in almost all cases. However, when compared against GPT-4, both Llama3 and MedLlama3 were significantly better for the 2021 collection. Additionally, 
MedLlama3 was statistically
better than Llama3 when no context was provided. On the other hand, 
the statistic tests confirmed that GPT-4 dominated Llama3/MedLlama3 in the 2022 collection.

The pairwise comparisons of ChatGPT vs d-003, Llama3 vs d-003, MedLlama3 vs d-003 and 
GPT-4 vs d-003 produced a higher frequency of statistically significant outcomes, but still, a majority of compared cases resulted in no significant differences.

Comparing the 0-shot results achieved
by LLMs (Figure \ref{fig:zeroshot}) with
those achieved by SEs (Figure \ref{fig:cum-accuracy-all}), we can observe
that, in general, LLMs produce higher
performance. 
Even the 
user behavior evaluation
(Figure \ref{fig:user-heuristic}), which simulates users skipping non-answers from SE results, yields effectiveness statistics that are inferior to those obtained with LLMs.
This suggests that the huge amounts of training data of the LLMs and their advanced reasoning capabilities are key advantages compared to the extraction
of responses from a few top-ranked search results. 

\begin{table}[t!]
\centering
\footnotesize

\begin{tabular}{l|cccc|cccc}
\multirow{2}{*}{prompt} & \multicolumn{4}{c|}{d002} & \multicolumn{4}{c}{d003}  \\
                        & \emph{0-shot} & \emph{1-shot} & \emph{2-shot} & \emph{3-shot} & \emph{0-shot} & \emph{1-shot} & \emph{2-shot} & \emph{3-shot}  \\
\hline
\hline
no-context              & 0.76   & 0.7   & \textbf{0.78}  & \textbf{0.78}  & 0.76   & \textbf{0.86} & \textbf{0.86}  & \textbf{0.86}   \\
non-expert              & 0.48   & \textbf{0.64}  & \textbf{0.74}  &\textbf{0.76}  & 0.72   & \textbf{0.82}  & \textbf{0.82}  & \textbf{0.82}    \\
expert                  & 0.68   & \textbf{0.74}*  & \textbf{0.76}*  & \textbf{0.78}*  & 0.72  & \textbf{0.82}*  & \textbf{0.84}*  & \textbf{0.84}*  \\
\hline
\hline
\end{tabular}

\vspace{0.25cm}
\begin{tabular}{l|cccc|cccc}
\centering
\multirow{2}{*}{prompt} & \multicolumn{4}{c|}{FT5} & \multicolumn{4}{c}{ChatGPT} \\
                        & \emph{0-shot} & \emph{1-shot} & \emph{2-shot} & \emph{3-shot} & \emph{0-shot} & \emph{1-shot} & \emph{2-shot} & \emph{3-shot}  \\
\hline
\hline
no-context          & 0.56   & \textbf{0.66}  & \textbf{0.64}  & \textbf{0.7}     & 0.76   & \textbf{0.82}  & \textbf{0.88}  & \textbf{0.84}  \\
non-expert       & 0.54   & \textbf{0.68}*  & \textbf{0.66}*  & \textbf{0.64}       & 0.8    & 0.8   & \textbf{0.88}  & \textbf{0.86}  \\
expert         & 0.74   & 0.68*  & 0.72  & 0.72          & 0.9    & 0.84  & 0.88  & 0.88  \\
\hline
\hline
\end{tabular}

\vspace{0.25cm}
\begin{tabular}{l|cccc|cccc}
\centering
\multirow{2}{*}{prompt} &  \multicolumn{4}{c|}{Llama3} & \multicolumn{4}{c}{GPT-4} \\
                         & \emph{0-shot} & \emph{1-shot} & \emph{2-shot} & \emph{3-shot} & \emph{0-shot} & \emph{1-shot} & \emph{2-shot} & \emph{3-shot}  \\
\hline
\hline
no-context                & 0.82   & \textbf{0.86}  & \textbf{0.84}  & 0.8 & 0.86   & 0.84  & 0.86  & 0.86  \\
non-expert                & 0.8   & 0.76  & 0.8  & 0.8  & 0.86   & 0.86  & \textbf{0.88}  & \textbf{0.88}  \\
expert                    & 0.8   & 0.76  & \textbf{0.82}  & 0.78   & 0.88   & 0.88  & \textbf{0.92}  & \textbf{0.9}   \\
\hline
\hline
\end{tabular}
\vspace{0.2cm}

\makebox[\textwidth][c]{
\begin{tabular}{l|cccc}
    \centering
\multirow{2}{*}{prompt} &  \multicolumn{4}{c}{MedLlama3} \\
                         & \emph{0-shot} & \emph{1-shot} & \emph{2-shot} & \emph{3-shot}   \\
\hline
\hline
no-context                & 0.78   & \textbf{0.8}  & 0.76  & 0.76   \\
non-expert                & 0.76   & \textbf{0.78}  & \textbf{0.78}  & \textbf{0.82}    \\
expert                    & 0.8   & \textbf{0.82}  & 0.8  & 0.8   \\
\hline
\hline
\end{tabular}
}
\vspace{0.2cm}
\caption{Few-shot experiments (with up to three in-context examples). Accuracy of each model and prompt. For each row, if a few-shot instance outperforms the 0-shot case then the few-shot case is bolded, and the symbol * marks those instances where the improvement was deemed 
as statistically significant
(McNemar's test, $\alpha=.05$).}
\label{tab:shot}
\end{table}

To examine the impact of in-context examples on LLMs, 
we made additional tests with
the health questions of TREC HM 2022. We sent each health question to the models preceded by one to three examples taken from TREC HM 2021. To this end, we selected three random pairs from the TREC HM 2021 dataset to serve as in-context examples and investigated their influence. Previous studies have indicated that a small set of in-context examples suffices for instructing the LLMs~\cite{liu2023pre}.

Table~\ref{tab:shot} reports the accuracy when providing a varying number of in-context examples (from 1-3). The impact of in-context examples varies significantly across different models. Specifically, 
FlanT5 and 
the two versions of GPT-3 (d-002 and d-003) showed a clearly positive effect
with in-context examples. For these three models, the inclusion of in-context cases resulted in statistically significant improvements. Conversely, the models that performed best in the zero-shot experiments did not experience any noticeable benefit from adding in-context examples. Regarding the type of prompts, the weakest prompts (``non expert'' and ``no context'') gained the most from in-context learning.
Furthermore, these experiments suggest that using one example is good enough
(adding more than one does not consistently improve performance).

In order to gain a deeper understanding of the performance of the LLMs, 
we scrutinized the instances in which all models were unsuccessful in delivering a correct response.
This analysis aimed to clarify the reasons for such low effectiveness. Specifically, we sent these ``low-performance questions'' again to the models, but this time, we did not set a limit to the token output. In this way, we could manually analyze the explanations provided and try to understand the reasoning behind the models. 
This examination was conducted with the most effective prompt (``expert'' variant) and with the ``no context'' variant, which arguably mirrors the type of interaction done by a lay user.

In the TREC HM 2020 collection, 
4\% of the questions had an incorrect answer
produced by all models using both prompts. 
The TREC HM 2021 collection had 2\% of
the questions assigned with wrong answers by all models
for the ``no context'' prompt and 6\% for the ``expert'' prompt.
Meanwhile, 
2\% of the questions from the TREC HM 2022 collection 
made that all LLMs 
failed for the ``no context'' prompt (and 4\% for the ``expert'' prompt). 
These figures confirm that, 
regardless of the LLM, 
some questions pose serious difficulties to the models.
 
After manually reviewing the models' completions, we were able to categorize the errors into a taxonomy that embodies the most frequent mistakes in health guidance, namely:

\begin{figure}[t!]
    \centering
    \includegraphics[width=0.55\textwidth]{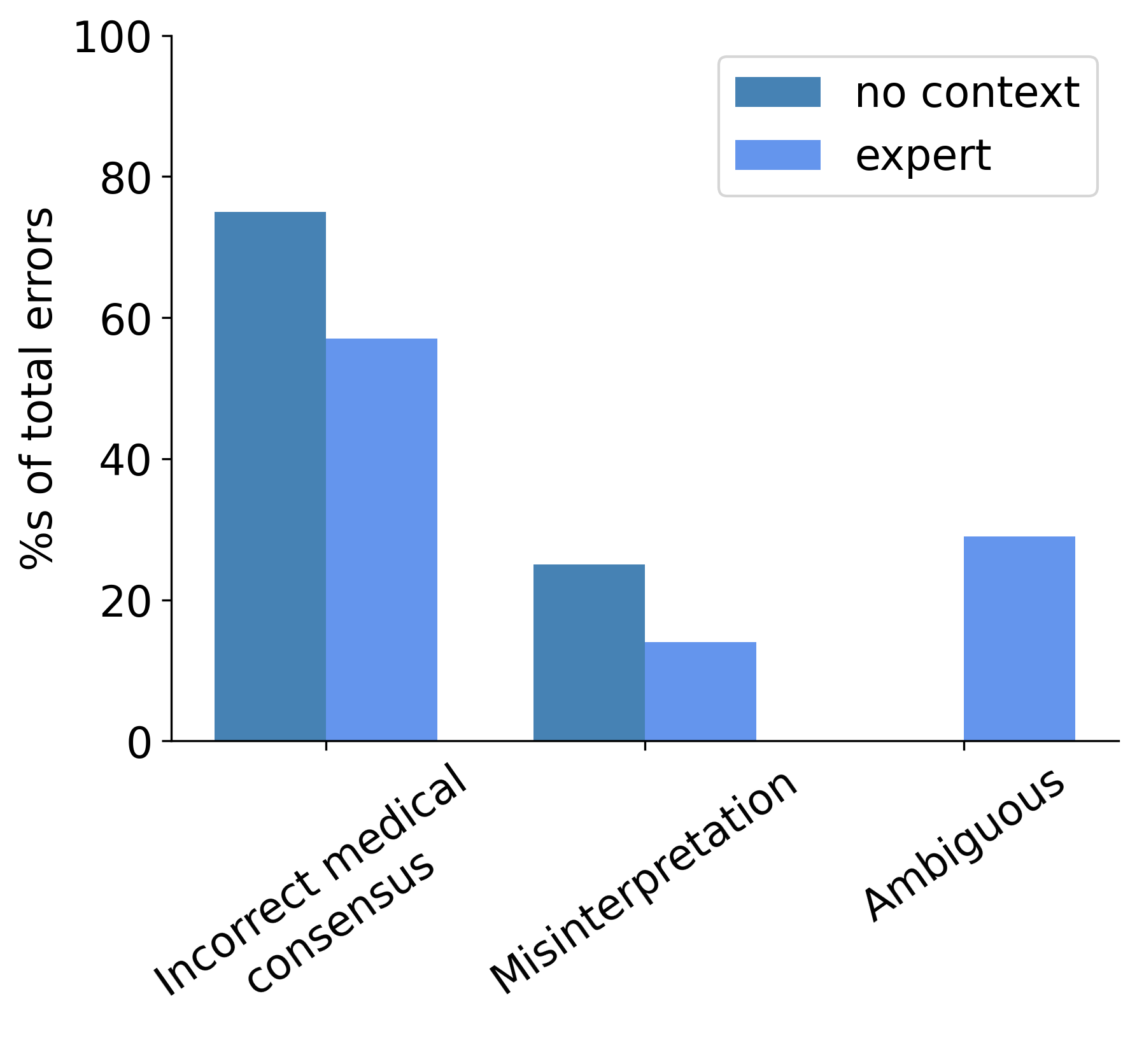}

  \caption{}
    \label{fig:errors}
\end{figure}

\begin{itemize}
    \item Incorrect understanding of prevailing medical consensus: Sometimes, the models provided responses that directly contradict the medical consensus. For example, 
    GPT-4 answered  ``Some research suggests that it may help in reducing symptoms associated with eczema...''    
    in response to ``Can evening primrose oil help treat 
    eczema?''. This answer clearly contradicts clinical evidence
    (see e.g., 
    \href{https://pmc.ncbi.nlm.nih.gov/articles/PMC292973/}{NIH does not support the use of evening pimrose oil to treat eczema}).
    \item Misinterpretation of the question: In this class of errors, the LLMs misunderstood the question. For instance, in response to ``Can bleach prevent COVID-19?'' the models produced completions such as ``No, bleach should not be ingested...''.  However, the correct interpretation of this inquiry is that the application of bleach for surface sterilization can indeed inhibit COVID-19.
    It is hardly conceivable that a human would interpret the question in the same fashion as the LLM.
    \item Ambiguous answer: This category includes responses where models did not generate a clear answer. These responses cannot be considered correct, but the LLM’s output could arguably be of value. For example, in reference to the potential effects of ankle braces to help heal an ankle fracture the model says that ``No, ankle braces alone do not heal an ankle fracture. They can, however, provide support, stability, and aid in managing pain during the healing process....''.
\end{itemize}

Figure~\ref{fig:errors} reports the percentage of each class of error for the different input conditions. 
The predominant category of error is the lack of knowledge about medical consensus. This is worrying because this is the most harmful type of mistake. Note also that the ``expert'' 
prompt had few
questions 
misinterpreted 
and few questions
whose answer contradicted
the medical consensus, but at the cost of providing more ambiguous answers. As future work, we would like to involve clinicians in reviewing the models' outputs and evaluating different dimensions, such as understandability or degree of correctness.

To further understand the relative merits of LLMs and SEs, we now analyze the behavior of the SEs with these difficult health questions. To that end, we calculated the average P@10 across all search engines for all questions and the average P@10 across all search engines for the difficult questions. 
The comparison of these two averages helps to understand whether or not the SEs also struggle with these topics: TREC HM 2020 topics (P@10 all: 0.53 vs P@10 difficult: 0.06), TREC HM 2021 topics (P@10 all: 0.50 vs P@10 difficult: 0.23), TREC HM 2022 topics (P@10 all: 0.60 vs P@10 difficult: 0.04).  
These figures suggest that the problem is not so much with LLMs as with the questions themselves, as search engines also return incorrect information in the top results for these difficult questions.

\subsection*{Retrieval-Augmented Large Language Models}

A stimulating line of research
consists of combining the potential of LLMs and SEs. Exploring the effect of including extracts from SE results in the LLMs' instructions is interesting. 
To reduce the monetary cost and computational load of these retrieval-augmented experiments, we decided to extract
the passages from a single engine (Google). The same passages
from Google 
were fed to five LLMs: text-davinci-002, ChatGPT, GPT-4, Llama3 and MedLlama3 under 
two prompting strategies (``no context'' and ``expert''). 

\begin{figure}[t!]
  \centering
    \includegraphics[width=\textwidth]{figures/Figure_6.png}
  

\caption{}
  \label{fig:rag-no-context}
\end{figure}

Figure~\ref{fig:rag-no-context} depicts the results of augmenting LLMs with each of the passages from Google's top 5 for the ``no context'' prompting strategy. For the 2020 questions, the LLMs do not seem to require these additional passages and, actually, four of the LLMs got worse results when presented with additional evidence.
For the two other datasets, the LLMs seem to improve their performance with textual evidence from the search engine. It must be noticed that in some cases (e.g., 2022), 
the provided passages make that the least sophisticated model, text-davinci-002, 
becomes comparable or even superior to more recent models, such as GPT-4 and reaches the performance of fine-tuned models in medical data. 
 We perceive this as an important outcome of our 
 study since it demonstrates that lighter models can achieve state-of-the-art performance when provided with additional evidence. In some cases, we can see a trend where the most sophisticated models only improve when provided with the first result. For example, this happened for MedLlama3 augmented with the top passage in the 2021 collection. 
 We estimated statistical significance using McNemar’s tests to compare 
 the augmented versions of each LLM 
 and its non-augmented counterpart. For the ``no context'' prompting strategy, two comparisons yielded significant differences (MedLlama3 no RA vs MedLlama3 + top 2 result -TREC HM 2022-, and ChatGPT no RA vs ChatGPT + top 1 result -TREC HM 2021-). 
 The rest of pairwise comparisons did not yield a significant outcome. This might be attributed to the relatively small size of the sample (number of queries in each collection).

On the other hand, Figure~\ref{fig:rag-expert} shows the behavior of the retrieval augmented language models with the ``expert'' prompt. Tendencies remain similar, but we can highlight a larger increase in performance for text-davinci-002 in the 2021 collection. Its performance grows from 0.36 up to values above 0.70 with retrieval augmentation. Again, we ran statistical tests and some comparisons declared statistical significance (ChatGPT no RA vs ChatGPT +  top 5 result -TREC HM 2021-,  GPT-4 no RA vs GPT-4 + top 1 result -TREC HM 2021-, Llama3 no RA vs Llama3 + top 2 result -TREC HM 2022-, Llama3 no RA vs Llama3 + top 3 result -TREC HM 2022-, and MedLlama3 no RA vs MedLlama3 + top 5 result -TREC HM 2022-).
We can conclude that there seems to be some positive signal from prompting the LLMs with SEs results, but the results are still inconclusive.

\begin{figure}[t!]
  \centering
  
      \includegraphics[width=\textwidth]{figures/Figure_7.png}


\caption{}
  \label{fig:rag-expert}
\end{figure}

We also evaluated the influence of augmenting with multiple passages. To that end, we ran experiments where the top 3 passages (extracted from Google's top 3 results) were concatenated and fed to the LLM as a single input. The 
results of this experiment were not very conclusive about the effect of injecting multiple passages. 
To further understand how the correctness of the injected passages influences the accuracy of the LLMs' responses,
we also analyzed the quality of the responses with varying
number of correct passages in the input. 
More specifically, for each passage obtained from Google
we extracted the answer (if any) that the passage gives to the question (using, again, the GPT3-based reading comprehension
module, see Methods). By contrasting these answers with the ground truth response of the health question, we can compute how many passages actually provide the LLM with a correct response. Since we injected three passages, we have four possible situations (0 correct passages out of 3, 1 correct passage out of 3, and so forth). Figure \ref{fig:rag-influence} plots the results of this analysis (for the sake of simplicity, we only ran this analysis for the expert prompt). As expected, 
accuracy improves as we feed the LLMs with higher proportions of correct passages. In fact, 
if all input passages are correct, then 
the LLMs hardly fail. 
In contrast, 
when all passages are incorrect, 
accuracy drops significantly, particularly in
the 2020 and 2021 collections. 
All models benefit from higher counts of
correct passages. This experiment confirms that the correctness of the injected passages is crucial to the performance of a health RAG system.

\begin{figure}[t!]
  \centering
  
      \includegraphics[width=\textwidth]{figures/Figure_8.png}


  \caption{}
  \label{fig:rag-influence}
\end{figure}

In the future, it would be interesting to explore other RAG variants 
and the interactions among LLM complexity, types of prompts, 
size and variety of retrieval results, and types of health questions.

\section*{Discussion}

Search engines are the 
classic information access
tools to retrieve content from online sources. The first goal of our study was to estimate their capacity to provide correct answers to binary health questions (RQ1). We have evaluated four popular SEs and observed that the percentage of correct answers found in the SERPs is in the range of 50\% to 70\%. 
These figures are low and might sound
highly concerning. 
However, this outcome is partially explained by the presence of many results that do not provide an answer. By focusing on the retrieved pages that provide an unequivocal answer to the reference health questions, we obtained much higher precision. Still, SE companies have room for improvement, as many 
top-ranked webpages do actually contain harmful health recommendations
(10\%-15\% of wrong answers). This is a natural consequence of the open and unmoderated nature of the web, and we encourage developers of search technologies to further advance in the removal of 
low-quality content from their indexes and SERPs. 

Regarding the presence of correct and incorrect answers over the ranked positions, the quality of the responses 
does not seem to decrease as we go deeper in the rankings
(at least for the top 20 results). Our results also show that Bing seems to be the most solid choice among the four SEs. As part of the analysis, we modeled two search user behaviors: lazy and diligent. 
We found out that lazy behavior, based on making decisions from the first observed response, is not inferior to a more diligent method based on acquiring and comparing three responses to the health questions.

Our next goal was determining whether LLMs are reliable for providing accurate medical answers (RQ2). Our results suggest
that, in general, the most capable 
LLMs generate better answers 
compared with those 
extracted from top webpages ranked by SEs. 
It seems that the extensive training
data of the LLMs, coupled with their superior reasoning abilities, offer significant advantages over extracting responses from a handful of top-ranked search results.
Among the largest models
(GPT-4, ChatGPT, LLama3, MedLLama3), we found no clear winner; and we observed poor performance from models such as FlanT5. 
Despite showing an overall good performance, there are still some barriers to adopting LLMs. For instance, under some circumstances, LLMs provide more than 30\% of incorrect results.

Another concerning outcome is that the quality of the LLMs' completions in response to health questions was highly sensitive to the input prompt
(RQ3). We found that some input prompts, which guide the models towards reputed sources, are much more effective than basic prompts (or prompts that give no context at all). However, lay users would hardly resort to sophisticated prompts or complex interactions with the LLMs. This suggests that the future adoption of LLMs to support QA in the health 
domain would need to wrap the user's questions into automatically extended contexts (or, alternatively, design health-oriented assistants that guide the AIs toward high-quality knowledge). 

We also conducted a thorough error analysis and demonstrated that LLMs made errors due to a lack of medical knowledge, even with the most sophisticated prompts. However, we also discovered that search engines also fail more in these ``low-performance questions", suggesting that the problem is not so much with the language models as with the questions themselves.

We also discovered that retrieval-augmented generation is promising for answering health questions (RQ4). We demonstrated that smaller LLMs reach the level of superior models when grounded 
in evidence provided by a search engine. 
This opens up the debate on whether it is worthwhile to persist in generating massive and computationally demanding models, or alternatively, we can direct our efforts towards leveraging lighter models enriched with search evidence. Moreover, we also demonstrated that the correctness of the passages used for augmenting the language model plays a critical role in its final performance.

In the literature, some teams focused their efforts on designing solutions that estimate the correctness of medical information. For instance, Vera ~\cite{pradeep2021vera} is a transformer-based ranker that achieved state-of-the-art performance in health misinformation detection tasks. This model was fine-tuned with assessments from TREC 2019's Decision Track and it achieved the best results in 
TREC 2022's Health Misinformation Track. This shared-data task fosters the development of systems prioritizing accurate and trustworthy documents over misinformation. 
However, systems like Vera take a health question and its correct response as input and then search for harmful or helpful documents. These technological tools could support, for example, moderation services for online platforms. However, the need for pairs of questions and correct responses represents a limitation. Thus, the creators of Vera also conducted research on how to automatically infer the correct response for a health question~\cite{pradeep}. Specifically, the authors evaluated two approaches: i) using LLMs under different settings (zero-, few-shot, and chain-of-thought), and ii) using the ranking produced by Vera and averaging the decisions extracted from the top 50 retrieved results. Their results suggest that the LLM-based approach (powered by GPT-4) outperformed the rest of the strategies. Our comparison of SEs and LLMs is related to this study, but we compared multiple commercial search engines (while Vera is not 
a tool that is 
widely available to the public). Furthermore, we do not only evaluate GPT models but also consider seven LLMs of different families. Our study, therefore, is reflective of the type of answers that the general public might get when interacting with popular information access tools. 
Additionally, we compared the answering capabilities of the LLMs with and without search results provided by the web search
engines. This is an aspect that has not received enough attention in health information seeking. We also evaluated the effectiveness of the answers provided as we go down in the ranked lists of the SERPs.

We are aware that our study presents some limitations. For instance, the automatic extraction of answers from specific passages for evaluating search engines can be prone to error. 
However, the passage retrieval and reading comprehension stages were based on state-of-the-art technologies that were effectively tested elsewhere (see Methods). We do not claim 
that the process is error-free but we are confident about the robustness of the trends found. 
In any case, in the near future we intend to further validate these findings by involving human evaluators to manually annotate the correctness of the search results. This is, however, a costly process that is not exempt from problems. 
Moreover, our current experiments did not consider the effect of personalization. Retrieval results are known to be dependent on multiple user factors (e.g., geolocalization or user preferences) and it will be important to study the relative quality of the medical responses taking into account geographical factors or other user-dependent variables.  
Related to this, it will be interesting
to study the relative adoption of LLMs and SEs 
by different groups of the population (e.g., educated/uneducated or wealthy/low-income).
Access to free vs. premium services provided by these platforms might represent a major factor in access to high-quality health-related information. We think
that this adoption should be scrutinized to ensure that the health gap between vulnerable people 
and other segments of the population does not enlarge.

Some research teams focused on other crucial aspects, such as readability or understandability. For example, Yan et al. \cite{yan2006} argued that search results should
be re-ranked by descending readability and proposed
several readability formulas for health and medical information retrieval. Related to this,
Zuccon and Koopman proposed an innovative method that integrates the notion of understandability into the evaluation of retrieval systems for consumer health 
search \cite{zuccon2014}.
In our study, we did not specifically analyze the understandability or readability of search results (or LLM's completions). These factors are known to affect user experience and we plan to extend our research to study the interaction between these two dimensions and health-related decisions.

About LLM evaluation, we are aware that using proprietary models poses the difficulty of reproducing these experiments. Their architectural design is unknown, the training data is not disclosed and, often, these are black-box models in constant evolution. 
However, we must acknowledge that these models are
currently being used by millions of people worldwide and it is of paramount importance to assess 
their answering capabilities in the health domain. Note that SEs are also closed systems whose core elements (e.g., ranking algorithms) are largely unknown and, additionally, their indexes undergo continuous updates. Our experiments and comparison, therefore, give a specific picture of LLMs and SEs at a certain point in time.  The forthcoming evolution of these systems will call for new trials and experiments to supplement the present study.
In any case, we facilitate the reproducibility of our experiments by providing the code (see Code availability), the generated outputs, 
and all the dates of execution of the experiments.

Another limitation affects the type of information needs. In these experiments, we restricted ourselves to binary question answering. 
This was a practical decision because we then were dealing with yes/no responses whose correctness can be automatically assessed.
This represents a valuable first step towards understanding the relative effectiveness of SEs and LLMs in the health domain.  
 In the near future, we would like to explore other classes of information needs that require more elaborate answers.

\section*{Methods}

\subsection*{Binary Question Answering}

In this research, we focus on estimating the ability of web search engines and LLMs to retrieve correct health responses. 
To that end, we selected
a binary health question answering (QA) task as our reference for evaluation. This type of information needs is prevalent, as many users usually search for specific advice, e.g., ``Can X (treatment) cure Y (disease)?''. 
This binary setting facilitates the quantitative assessment of the system's responses. Automatic systems must provide the correct answer to these health questions and, as shown below, the target or reference response comes from the established 
medical consensus for each search topic.

\begin{figure}
    \centering
    \includegraphics[width=0.7\linewidth]{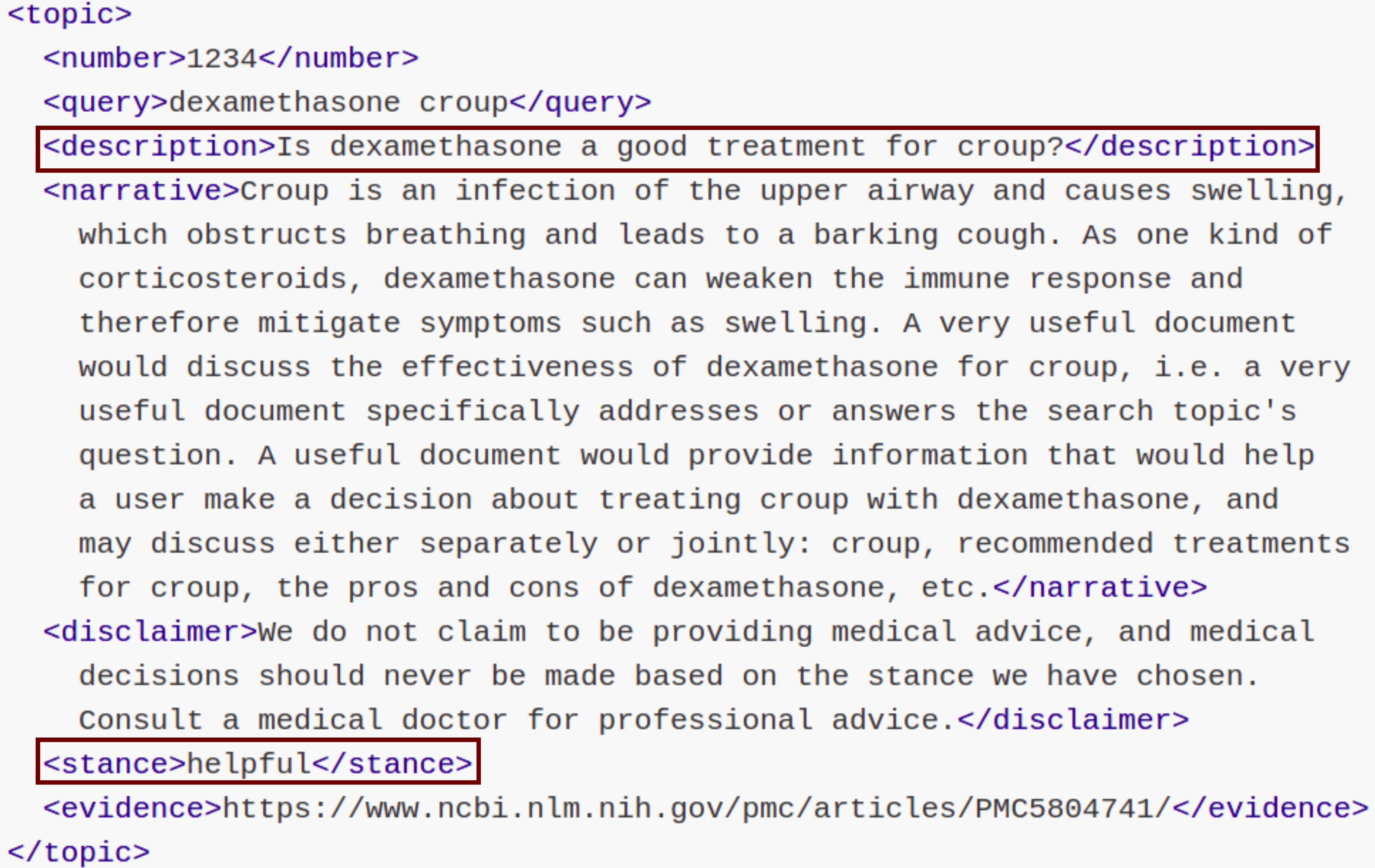}
      \caption{}
    \label{fig:topic}
\end{figure}

To obtain a solid set of health-related
search topics we leveraged the data 
created under the TREC Health Misinformation Track.
This is a three-year shared-data task that fostered the creation of systems adept at detecting false health information, thereby empowering individuals to make health-related choices grounded on reliable and factual information~\cite{clarke2020overview,clarke2021overview}. These datasets contain health-related queries posed as questions (for instance, ``Does wearing masks prevent COVID-19?'') and their correct responses (yes/no).
Each topic represents a searcher who is looking for information that is useful for making a ``yes'' or ``no'' decision regarding a health-related information need. The binary ground truth field
represents the current 
understanding of contemporary medical practice (compiled by the 
organizers of the tasks in the process
of constructing the dataset). Figure~\ref{fig:topic} shows an example of a topic (the question is stored in the description field while the binary response is stored in the stance field, encoded as ``helpful'' or ``unhelpful'').

The TREC HM 2020 dataset was restricted to questions related to COVID-19, while topics from TREC HM 2021 and 2022 collections covered a wide range of health topics. 
The 2020 questions were disclosed in the middle of 2020, which raises the possibility that they could have been included in the pre-training phase of some LLMs. Such inclusion could give them an advantage over generative models trained earlier. 
The 2021 dataset was made available in 
the middle of 2021 and, therefore, it could have been used for training models such as 
GPT-4 or ChatGPT  
(but GPT-3, instead, ended its training earlier). The topics of 2022, on the other hand, were released after all the LLMs had been created. Therefore, this configuration of health topics shapes an assorted evaluation with questions created and released at different dates. 

Summing up, the selection of health topics
integrates a diverse set of binary health-related questions, which pose different degrees of difficulty related to the exposure of the models to such data and to the specificity of the information needs.

\subsection*{Search Engines}

To evaluate the responses provided by SEs, 
each health question was submitted to four well-known search engines: Google, Bing, Yahoo, and Duckduckgo. We 
used a scraping tool (see Code Availability) on the organic search results to collect the top retrieved webpages. 
We gathered the top 20 ranked entries since
users rarely go beyond the second page of results, and thus, 
the extraction focuses on webpages that 
have some chance of being inspected by a standard user. Then, we obtained the raw content of the webpage, converted it 
to workable plain text, and split it into passages. To obtain 
on-topic passages, the health question and each passage were embedded into a vectorial contextual representation using MonoT5 ~\cite{pradeep2023pygaggle}, a highly effective model fine-tuned for passage retrieval.
This constitutes a well-established approach for the automatic extraction of excerpts answering user queries \cite{nogueira2019passage}. Rosa and colleagues \cite{rosa2022no} demonstrated the effectiveness of this model for several retrieval tasks. 

\begin{figure}[t!]
   \centering
   \makebox[\linewidth]{\includegraphics[width=0.9\linewidth]{figures/Figure_10.png}} 

  \caption{}
   \label{fig:se-architecture}
\end{figure}

After the passages were chosen, 
it was necessary to ascertain whether they provided an affirmative, negative, or non-responsive answer to the health question.
To that end, we exploited GPT-3 model's capabilities to perform reading comprehension. 
Brown et al. \cite{brown2020language} evaluated the ability of this model in different reading comprehension settings. The model achieved remarkable results, for example, F1 of 85\% in the CoQA dataset. Dijkstra and others \cite{dijkstra2022reading} also demonstrated this model's ability to perform reading comprehension tasks in the educational domain. In our case, we defined a prompt that included the passage and health question, and specifically tasked 
the model with answering 
the question based on the provided passage
(without resorting to its internal knowledge).
The following prompt template was used: 
``$<$Passage$>$. Based on the previous text, answer `yes', `no' or `no answer provided' to the following question: $<$health question$>$''. 
This estimation process arguably produces the sequence of answers that a search engine user would obtain from inspecting
the SE results. 
Figure~\ref{fig:se-architecture} illustrates the whole process. 

Note that 
a retrieval result might not provide an answer to the question. 
In our accounting of correct answers across ranking positions, these cases are recorded as failures, 
as they do not respond with the correct answer. It is important, though, 
to bear in mind that there is an important distinction between a non-answer
and an incorrect answer. We will further delve into this issue in the future, but in the current study, we focused on analyzing the relative trends of correct responses.

Our interest here is not only to analyze the overall effectiveness of web search systems but also to study the quality of the responses as we go deeper into the ranked lists. 
To that end, we considered two models of user behavior that simulate alternative forms of inspection of the retrieved results, see Figure~\ref{fig:user-models}. 
The lazy user model represents a user who stops inspecting results when presented with the first entry that gives a yes/no response to the user's question. Note that on-topic passages might not respond to the user's question, and thus, the lazy user model does not necessarily stop at the top 1 entry. 
This user, therefore, sticks to the
first answer found and does not spend time searching for contrasting evidence. 
We therefore assume that this user believes the answer provided by the passage. This class of user behavior (blindly believing the first response found)
is clearly suboptimal, given the high presence of misinformation on the web.

\begin{figure}[t!]
   \centering
   \includegraphics[width=0.9\linewidth]{figures/Figure_11.png}
   \caption{}
   \label{fig:user-models}
\end{figure}

The second model, referred to as diligent user model, represents a user who traverses the ranking from the top position and stops after finding three responses. We assume that this user makes a decision about his health question based on majority voting from the three provided responses. 

Under these models, the user's decision is the binary resolution the user takes after seeing the first answer (lazy model) or after seeing three answers (diligent model). It might be helpful or harmful, depending on the correctness/incorrectness of the documents inspected.

Web users are known to be reluctant to explore many search results, and these two models represent two natural forms of exploration in the quest for the response to the health question. For future work, we leave the study of additional user models, including click models for web search \cite{chuklin2015click}.

\subsection*{Large Language Models}

To assess the LLMs, 
we submitted the same health question 
to several generative models
and obtained their completions. 
We worked under different settings, including zero- and few-shot strategies (providing examples of correct responses to other health questions), and different types of prompts. 

Specifically, we evaluated seven models of different nature and architecture (closed and open source), including 
a model fine-tuned to the clinical domain. The fine-tuned model helps to test the influence of in-domain data in accurately answering health questions. The models included in our comparative study are:

\begin{itemize}
\item GPT-3, also known as text-davinci-002 (d-002). This model has 
175 billion parameters and a decoder-only structure.
It was extensively trained with corpora
from diverse sources (including a full Wikipedia corpus). The training information goes up to June 2021. 

\item text-davinci-003 (d-003), the subsequent iteration of GPT models, built upon its predecessor by incorporating InstructGPT methodologies~\cite{ouyang2022training}. This model has been built through reinforcement learning with human feedback (RLHF) and has the same training data timeline as d-002.

\item ChatGPT represents an evolution of OpenAI models towards a more dialogue-oriented and user-friendly behavior. Its knowledge cutoff goes until September 2021. 
We experimented with the 
gpt-3.5-turbo version (from June 2023).

\item GPT-4, another conversational agent 
by OpenAI
that represents a significant leap forward, outperforming ChatGPT in complex tasks that demand human-like reasoning, such as solving academic exercises~\cite{gpt4}. The training 
examples of GPT-4 were collected until September 2021.

\item Flan T5 (FT5), a sequence-to-sequence model from Google. This model underwent fine-tuning with a diverse array of instruction-based datasets~\cite{longpre2023flan}. The data for this model was sourced from the ``Flan 2022'' open-source repository, which includes online 
content collected until the end of 2022. For our experimentation, we worked with flan-t5-xl.

\item  Llama3, the most recent language model developed by Meta AI. Llama3's pretraining data covers until the end of 2023. It also underwent several post-training rounds, aligning the model with human feedback. Each post-training round included supervised fine-tuning and preference optimization through reinforcement learning. For these experiments, we used the version named llama3.1-8b-instruct.

\item MedLlama3, a fine-tuned Llama3 with medical instruction data~\cite{christophe2024med42}. It was specially tuned for user alignment in healthcare settings. We used Med42-Llama3-8B version, released in August 2024.

\end{itemize}

The four OpenAI models were tested through its official Python API, while 
Llama3, MedLlama3 
and Flan T5 
were executed through the Ollama and
HuggingFace APIs, respectively (see Code Availability for further details). 
We fixed the temperature of the 
models to 0, with the intention of reducing the randomness or creativity of the completions. 

A pivotal part of this research consists of determining the effectiveness of these models for answering health questions under different input conditions or contexts. 
Having in mind that online users are known to be reluctant to have complex interactions with automated systems, we first tested the LLMs' effectiveness when responding
to non-expert individuals who 
send the question and give little or no context at all:

\begin{itemize}
\item no-context prompt: A context formed only with the health question, i.e. ``Can Vitamin D cure COVID-19?''.
\item non-expert prompt: The text ``I am a non-expert user searching for medical advice online'' is added before the health question. This prompt intends to mirror a lay user seeking medical counsel.  
\end{itemize}

As a next step, we tested more elaborate prompts and
we also assessed the
influence of including some in-context examples. 
While it may be uncommon for a typical user to adopt these sophisticated strategies,
they nonetheless offer valuable insights into understanding and leveraging the inherent knowledge of the models:

\begin{itemize}
    \item expert prompt: The text ``We are a committee of leading scientific experts and medical doctors reviewing the latest and highest quality of research from PubMED. For each question, we have chosen an answer, either `yes' or `no', based on our best understanding of current medical practice and literature.'' followed by the corresponding medical question. 
    These contextual instructions were produced by 
    the research team
    from Waterloo University
    for their TREC 2022 HM experiments~\cite{pradeep}. The core idea is to guide the LLM towards renowned sources.
\end{itemize}

To automatically record responses, we forced the models to answer only with ``yes'' or ``no'' tokens. 
In contrast to the scenario with search engines, there can, therefore, be no unanswered questions here.
More complex prompt engineering alternatives, like Chain-of-Thought (CoT)~\cite{cot2022}, were left for future work.

For the few-shot experiments, we selected three random pairs from the TREC HM 2021 dataset. The selected question-answer pairs were: (``Does yoga help manage asthma?'', ``Yes''), (``Will wearing an ankle brace help heal achilles tendonitis?'', ``No''), and (``Is starving a fever effective?'', ``No'').

In this study, we also performed an ``online'' retrieval augmented generation experiment~\cite{asai2023retrieval}. There are several approaches for augmenting generative language models with retrieved evidence. 
Following standard practice, 
we injected textual chunks in the input layer of the LLM \cite{guu2020retrieval}. In essence, we prompted different large language models with passages extracted from the top entries of the SERPs.
In this way, our experimental setup allows the comparison of three classes of outputs (SE alone, LLM alone,
and a RAG variant that injects real-time SE results into the input of the LLM). 
For these RAG experiments, we fed some LLMs with on-topic passages obtained from the search results produced by Google. We injected passages from Google's top 5 results
and we clearly instructed the LLM to compare the provided evidence with its internal knowledge prior to providing a definitive answer. The following prompt was used for RAG: ``Provide an answer to the question using the provided evidence and contrasting it with your internal knowledge. Evidence: $<$Evidence from search engine$>$. Question: $<$query$>$. Your answer:''. The extraction of on-topic passages from Google's top 5 results was also supported by MonoT5.

\section*{Data availability}

All data used in this study is public and it can be found in the Text Retrieval Conference (TREC) repository (\url{https://trec.nist.gov/data.html}).

\section*{Code availability}

All code used in this experimentation is openly published under a GPLv3 License ({\url{https://github.com/MarcosFP97/information-seeking-health}). Special mention to the open source scraping tool used to evaluate the search engines (\url{https://github.com/tasos-py/Search-Engines-Scraper}). 

The four OpenAI models were tested through its official Python API (\url{https://openai.com/blog/openai-api}), while 
Llama3 (\url{https://ollama.com/library/llama3.1:8b-instruct-q8_0}, a quantized version of 8 bits of the model to run on a CPU), MedLlama3 (\url{https://ollama.com/thewindmom/llama3-med42-8b}, quantized version, 8 bits) 
and Flan T5 (\url{https://huggingface.co/google/flan-t5-xl}) 
were executed through the Ollama and
HuggingFace APIs, respectively. 

\section*{Acknowledgements}

The authors thank the financial support supplied by 
 the Xunta de Galicia - Consellería de Cultura, Educación, Formación Profesional e Universidades  (ED431G 2023/04, ED431C 2022/19) and
the ERDF, which acknowledges the CiTIUS Research Center in Intelligent Technologies of
the USC as a Research Center of the Galician
University System. 
They  also thank 
the financial support obtained from i) project PID2022-137061OB-C22 
(Ministerio de Ciencia e Innovación, Agencia Estatal de Investigación, 
Proyectos de Generación de Conocimiento; suppported by the European Regional Development Fund) 
and ii) project PLEC2021-007662 (MCIN/AEI/10.13039/501100011033, Ministerio de
Ciencia e Innovación, Agencia Estatal de Investigación, Plan de Recuperación,
Transformación y Resiliencia, Unión Europea-Next Generation EU).

\bibliography{sn-bibliography}

\begin{thebibliography}{10}
\expandafter\ifx\csname url\endcsname\relax
  \def\url#1{\burl{#1}}\fi
\expandafter\ifx\csname urlprefix\endcsname\relax\def\urlprefix{URL }\fi
\providecommand{\bibinfo}[2]{#2}
\providecommand{\eprint}[2][]{\url{#2}}
\providecommand{\doi}[1]{\url{https://doi.org/#1}}
\bibcommenthead

\bibitem{longpre2023flan}
\bibinfo{author}{Longpre, S.} \emph{et~al.}
\newblock \bibinfo{title}{The flan collection: Designing data and methods for effective instruction tuning}.
\newblock \emph{\bibinfo{journal}{arXiv preprint arXiv:2301.13688}}  (\bibinfo{year}{2023}).

\bibitem{bubeck2023sparks}
\bibinfo{author}{Bubeck, S.} \emph{et~al.}
\newblock \bibinfo{title}{Sparks of artificial general intelligence: Early experiments with {GPT-4}}.
\newblock \emph{\bibinfo{journal}{arXiv preprint arXiv:2303.12712}}  (\bibinfo{year}{2023}).

\bibitem{touvron2023llama}
\bibinfo{author}{Touvron, H.} \emph{et~al.}
\newblock \bibinfo{title}{Llama 2: Open foundation and fine-tuned chat models}.
\newblock \emph{\bibinfo{journal}{arXiv preprint arXiv:2307.09288}}  (\bibinfo{year}{2023}).

\bibitem{mao2023large}
\bibinfo{author}{Mao, K.}, \bibinfo{author}{Dou, Z.}, \bibinfo{author}{Chen, H.}, \bibinfo{author}{Mo, F.} \& \bibinfo{author}{Qian, H.}
\newblock \bibinfo{title}{Large language models know your contextual search intent: A prompting framework for conversational search}.
\newblock \emph{\bibinfo{journal}{arXiv preprint arXiv:2303.06573}}  (\bibinfo{year}{2023}).

\bibitem{friedman2023leveraging}
\bibinfo{author}{Friedman, L.} \emph{et~al.}
\newblock \bibinfo{title}{Leveraging large language models in conversational recommender systems}.
\newblock \emph{\bibinfo{journal}{arXiv preprint arXiv:2305.07961}}  (\bibinfo{year}{2023}).

\bibitem{polak2023extracting}
\bibinfo{author}{Polak, M.~P.} \& \bibinfo{author}{Morgan, D.}
\newblock \bibinfo{title}{Extracting accurate materials data from research papers with conversational language models and prompt engineering--example of chatgpt}.
\newblock \emph{\bibinfo{journal}{arXiv preprint arXiv:2303.05352}}  (\bibinfo{year}{2023}).

\bibitem{o2022massive}
\bibinfo{author}{O’Leary, D.~E.}
\newblock \bibinfo{title}{Massive data language models and conversational artificial intelligence: Emerging issues}.
\newblock \emph{\bibinfo{journal}{Intelligent Systems in Accounting, Finance and Management}} \textbf{\bibinfo{volume}{29}}, \bibinfo{pages}{182--198} (\bibinfo{year}{2022}).

\bibitem{hersh2024search}
\bibinfo{author}{Hersh, W.}
\newblock \bibinfo{title}{Search still matters: information retrieval in the era of generative {AI}}.
\newblock \emph{\bibinfo{journal}{Journal of the American Medical Informatics Association}} \textbf{\bibinfo{volume}{31}}, \bibinfo{pages}{2159--2161} (\bibinfo{year}{2024}).

\bibitem{wu2024investigating}
\bibinfo{author}{Wu, B.}, \bibinfo{author}{Liu, Q.~B.}, \bibinfo{author}{Guo, X.} \& \bibinfo{author}{Yang, C.}
\newblock \bibinfo{title}{Investigating patients' adoption of online medical advice}.
\newblock \emph{\bibinfo{journal}{Decision Support Systems}} \textbf{\bibinfo{volume}{176}}, \bibinfo{pages}{114050} (\bibinfo{year}{2024}).

\bibitem{sivarajkumar2024clinical}
\bibinfo{author}{Sivarajkumar, S.} \emph{et~al.}
\newblock \bibinfo{title}{Clinical information retrieval: A literature review}.
\newblock \emph{\bibinfo{journal}{Journal of Healthcare Informatics Research}} \textbf{\bibinfo{volume}{8}}, \bibinfo{pages}{313--352} (\bibinfo{year}{2024}).

\bibitem{wang2012using}
\bibinfo{author}{Wang, L.} \emph{et~al.}
\newblock \bibinfo{title}{Using internet search engines to obtain medical information: a comparative study}.
\newblock \emph{\bibinfo{journal}{Journal of medical Internet research}} \textbf{\bibinfo{volume}{14}}, \bibinfo{pages}{e74} (\bibinfo{year}{2012}).

\bibitem{zuccon2015}
\bibinfo{author}{Zuccon, G.}, \bibinfo{author}{Koopman, B.} \& \bibinfo{author}{Palotti, J.}
\newblock \bibinfo{title}{Diagnose this if you can: on the effectiveness of search engines in finding medical self-diagnosis information}.
\newblock \emph{\bibinfo{journal}{Advances in Information Retrieval}} \textbf{\bibinfo{volume}{9022}}, \bibinfo{pages}{562--567} (\bibinfo{year}{2015}).

\bibitem{jiang2020can}
\bibinfo{author}{Jiang, Z.}, \bibinfo{author}{Xu, F.~F.}, \bibinfo{author}{Araki, J.} \& \bibinfo{author}{Neubig, G.}
\newblock \bibinfo{title}{How can we know what language models know?}
\newblock \emph{\bibinfo{journal}{Transactions of the Association for Computational Linguistics}} \textbf{\bibinfo{volume}{8}}, \bibinfo{pages}{423--438} (\bibinfo{year}{2020}).

\bibitem{liang2022holistic}
\bibinfo{author}{Liang, P.} \emph{et~al.}
\newblock \bibinfo{title}{Holistic evaluation of language models}.
\newblock \emph{\bibinfo{journal}{arXiv preprint arXiv:2211.09110}}  (\bibinfo{year}{2022}).

\bibitem{chang2024survey}
\bibinfo{author}{Chang, Y.} \emph{et~al.}
\newblock \bibinfo{title}{A survey on evaluation of large language models}.
\newblock \emph{\bibinfo{journal}{ACM Transactions on Intelligent Systems and Technology}} \textbf{\bibinfo{volume}{15}}, \bibinfo{pages}{1--45} (\bibinfo{year}{2024}).

\bibitem{chervenak2023promise}
\bibinfo{author}{Chervenak, J.}, \bibinfo{author}{Lieman, H.}, \bibinfo{author}{Blanco-Breindel, M.} \& \bibinfo{author}{Jindal, S.}
\newblock \bibinfo{title}{The promise and peril of using a large language model to obtain clinical information: {C}hat{GPT} performs strongly as a fertility counseling tool with limitations}.
\newblock \emph{\bibinfo{journal}{Fertility and Sterility}} \textbf{\bibinfo{volume}{120}}, \bibinfo{pages}{575--583} (\bibinfo{year}{2023}).

\bibitem{duong2023analysis}
\bibinfo{author}{Duong, D.} \& \bibinfo{author}{Solomon, B.~D.}
\newblock \bibinfo{title}{Analysis of large-language model versus human performance for genetics questions}.
\newblock \emph{\bibinfo{journal}{European Journal of Human Genetics}} \textbf{\bibinfo{volume}{32}}, \bibinfo{pages}{466--468} (\bibinfo{year}{2023}).

\bibitem{holmes2023evaluating}
\bibinfo{author}{Holmes, J.} \emph{et~al.}
\newblock \bibinfo{title}{Evaluating large language models on a highly-specialized topic, radiation oncology physics}.
\newblock \emph{\bibinfo{journal}{arXiv preprint arXiv:2304.01938}}  (\bibinfo{year}{2023}).

\bibitem{elgedawy2024dynamic}
\bibinfo{author}{Elgedawy, R.}, \bibinfo{author}{Srinivasan, S.} \& \bibinfo{author}{Danciu, I.}
\newblock \bibinfo{title}{Dynamic {Q\&A} of clinical documents with large language models}.
\newblock \emph{\bibinfo{journal}{arXiv preprint arXiv:2401.10733}}  (\bibinfo{year}{2024}).

\bibitem{kim2024assessing}
\bibinfo{author}{Kim, H.-W.}, \bibinfo{author}{Shin, D.-H.}, \bibinfo{author}{Kim, J.}, \bibinfo{author}{Lee, G.-H.} \& \bibinfo{author}{Cho, J.~W.}
\newblock \bibinfo{title}{Assessing the performance of chatgpt's responses to questions related to epilepsy: A cross-sectional study on natural language processing and medical information retrieval}.
\newblock \emph{\bibinfo{journal}{Seizure: European Journal of Epilepsy}} \textbf{\bibinfo{volume}{114}}, \bibinfo{pages}{1--8} (\bibinfo{year}{2024}).

\bibitem{kim2024large}
\bibinfo{author}{Kim, J.} \emph{et~al.}
\newblock \bibinfo{title}{Large language models outperform mental and medical health care professionals in identifying obsessive-compulsive disorder}.
\newblock \emph{\bibinfo{journal}{NPJ Digital Medicine}} \textbf{\bibinfo{volume}{7}}, \bibinfo{pages}{193} (\bibinfo{year}{2024}).

\bibitem{jahan2023evaluation}
\bibinfo{author}{Jahan, I.}, \bibinfo{author}{Laskar, M. T.~R.}, \bibinfo{author}{Peng, C.} \& \bibinfo{author}{Huang, J.}
\newblock \bibinfo{title}{Evaluation of {ChatGPT} on biomedical tasks: A zero-shot comparison with fine-tuned generative transformers}.
\newblock \emph{\bibinfo{journal}{arXiv preprint arXiv:2306.04504}}  (\bibinfo{year}{2023}).

\bibitem{hamidi2023evaluation}
\bibinfo{author}{Hamidi, A.} \& \bibinfo{author}{Roberts, K.}
\newblock \bibinfo{title}{Evaluation of {AI} chatbots for patient-specific {EHR} questions}.
\newblock \emph{\bibinfo{journal}{arXiv preprint arXiv:2306.02549}}  (\bibinfo{year}{2023}).

\bibitem{samaan2023assessing}
\bibinfo{author}{Samaan, J.~S.} \emph{et~al.}
\newblock \bibinfo{title}{Assessing the accuracy of responses by the language model chatgpt to questions regarding bariatric surgery}.
\newblock \emph{\bibinfo{journal}{Obesity surgery}} \textbf{\bibinfo{volume}{33}}, \bibinfo{pages}{1790--1796} (\bibinfo{year}{2023}).

\bibitem{tang2023evaluating}
\bibinfo{author}{Tang, L.} \emph{et~al.}
\newblock \bibinfo{title}{Evaluating large language models on medical evidence summarization}.
\newblock \emph{\bibinfo{journal}{NPJ digital medicine}} \textbf{\bibinfo{volume}{6}}, \bibinfo{pages}{158} (\bibinfo{year}{2023}).

\bibitem{johnson2023assessing}
\bibinfo{author}{Johnson, D.} \emph{et~al.}
\newblock \bibinfo{title}{Assessing the accuracy and reliability of ai-generated medical responses: an evaluation of the chat-gpt model}.
\newblock \emph{\bibinfo{journal}{Preprint National Library of Medicine}}  (\bibinfo{year}{2023}).

\bibitem{thirunavukarasu2023trialling}
\bibinfo{author}{Thirunavukarasu, A.~J.} \emph{et~al.}
\newblock \bibinfo{title}{Trialling a large language model (chatgpt) in general practice with the applied knowledge test: observational study demonstrating opportunities and limitations in primary care}.
\newblock \emph{\bibinfo{journal}{JMIR Medical Education}} \textbf{\bibinfo{volume}{9}}, \bibinfo{pages}{e46599} (\bibinfo{year}{2023}).

\bibitem{kusa-etal-2023-dr}
\bibinfo{author}{Kusa, W.}, \bibinfo{author}{Mosca, E.} \& \bibinfo{author}{Lipani, A.}
\newblock \bibinfo{title}{``{Dr {LLM}, what do I have?}'': The impact of user beliefs and prompt formulation on health diagnoses}.
\newblock \emph{\bibinfo{journal}{Proceedings of the Third Workshop on NLP for Medical Conversations}} \bibinfo{pages}{13\textendash19} (\bibinfo{year}{2023}).

\bibitem{caramancion2024large}
\bibinfo{author}{Caramancion, K.~M.}
\newblock \bibinfo{title}{Large language models vs. search engines: Evaluating user preferences across varied information retrieval scenarios}.
\newblock \emph{\bibinfo{journal}{arXiv preprint arXiv:2401.05761}}  (\bibinfo{year}{2024}).

\bibitem{oeding2024chatgpt}
\bibinfo{author}{Oeding, J.~F.} \emph{et~al.}
\newblock \bibinfo{title}{Chat{GPT}-4 performs clinical information retrieval tasks utilizing consistently more trustworthy resources than does {G}oogle {S}earch for queries concerning the {L}atarjet procedure}.
\newblock \emph{\bibinfo{journal}{Arthroscopy: The Journal of Arthroscopic \& Related Surgery}}  (\bibinfo{year}{2024}).

\bibitem{li2023chatdoctor}
\bibinfo{author}{Li, Y.} \emph{et~al.}
\newblock \bibinfo{title}{{ChatDoctor}: A medical chat model fine-tuned on a large language model {Meta-AI (LLaMA)} using medical domain knowledge}.
\newblock \emph{\bibinfo{journal}{Cureus}} \textbf{\bibinfo{volume}{15}} (\bibinfo{year}{2023}).

\bibitem{koopman2023dr}
\bibinfo{author}{Koopman, B.} \& \bibinfo{author}{Zuccon, G.}
\newblock \bibinfo{title}{Dr {ChatGPT} tell me what i want to hear: How different prompts impact health answer correctness}.
\newblock \emph{\bibinfo{journal}{Proceedings of the 2023 Conference on Empirical Methods in Natural Language Processing}} \bibinfo{pages}{15012--15022} (\bibinfo{year}{2023}).

\bibitem{xiong2024benchmarking}
\bibinfo{author}{Xiong, G.}, \bibinfo{author}{Jin, Q.}, \bibinfo{author}{Lu, Z.} \& \bibinfo{author}{Zhang, A.}
\newblock \bibinfo{title}{Benchmarking retrieval-augmented generation for medicine}.
\newblock \emph{\bibinfo{journal}{arXiv preprint arXiv:2402.13178}}  (\bibinfo{year}{2024}).

\bibitem{brown2020language}
\bibinfo{author}{Brown, T.} \emph{et~al.}
\newblock \bibinfo{title}{Language models are few-shot learners}.
\newblock \emph{\bibinfo{journal}{Advances in neural information processing systems}} \textbf{\bibinfo{volume}{33}}, \bibinfo{pages}{1877--1901} (\bibinfo{year}{2020}).

\bibitem{liu2023pre}
\bibinfo{author}{Liu, P.} \emph{et~al.}
\newblock \bibinfo{title}{Pre-train, prompt, and predict: A systematic survey of prompting methods in natural language processing}.
\newblock \emph{\bibinfo{journal}{ACM Computing Surveys}} \textbf{\bibinfo{volume}{55}}, \bibinfo{pages}{1--35} (\bibinfo{year}{2023}).

\bibitem{pradeep2021vera}
\bibinfo{author}{Pradeep, R.}, \bibinfo{author}{Ma, X.}, \bibinfo{author}{Nogueira, R.} \& \bibinfo{author}{Lin, J.}
\newblock \bibinfo{title}{Vera: Prediction techniques for reducing harmful misinformation in consumer health search}.
\newblock \emph{\bibinfo{journal}{Proceedings of the 44th International ACM SIGIR Conference on Research and Development in Information Retrieval}} \bibinfo{pages}{2066--2070} (\bibinfo{year}{2021}).

\bibitem{pradeep}
\bibinfo{author}{Pradeep, R.} \& \bibinfo{author}{Lin, J.}
\newblock \bibinfo{title}{Towards automated end-to-end health misinformation free search with a large language model}.
\newblock \emph{\bibinfo{journal}{Advances in Information Retrieval}} \bibinfo{pages}{78--86} (\bibinfo{year}{2024}).

\bibitem{yan2006}
\bibinfo{author}{Yan, X.}, \bibinfo{author}{Song, D.} \& \bibinfo{author}{Li, X.}
\newblock \bibinfo{title}{Concept-based document readability in domain specific information retrieval}.
\newblock \emph{\bibinfo{journal}{Proceedings of the 15th ACM International Conference on Information and Knowledge Management}} \bibinfo{pages}{540–549} (\bibinfo{year}{2006}).
\newblock \urlprefix\url{https://doi.org/10.1145/1183614.1183692}.

\bibitem{zuccon2014}
\bibinfo{author}{Zuccon, G.} \& \bibinfo{author}{Koopman, B.}
\newblock \bibinfo{title}{Integrating understandability in the evaluation of consumer health search engines}.
\newblock \emph{\bibinfo{journal}{CEUR Workshop Proceedings}} \textbf{\bibinfo{volume}{1276}}, \bibinfo{pages}{32--35} (\bibinfo{year}{2014}).

\bibitem{clarke2020overview}
\bibinfo{author}{Clarke, C.}, \bibinfo{author}{Maistro, M.}, \bibinfo{author}{Smucker, M.} \& \bibinfo{author}{Zuccon, G.}
\newblock \bibinfo{title}{{Overview of the TREC 2020 Health Misinformation Track}}.
\newblock \emph{\bibinfo{journal}{Proceedings of the Twenty-Nine Text REtrieval Conference, TREC}} \bibinfo{pages}{16--19} (\bibinfo{year}{2020}).

\bibitem{clarke2021overview}
\bibinfo{author}{Clarke, C.}, \bibinfo{author}{Maistro, M.} \& \bibinfo{author}{Smucker, M.}
\newblock \bibinfo{title}{{Overview of the TREC 2021 Health Misinformation Track}}.
\newblock \emph{\bibinfo{journal}{Proceedings of the Thirtieth Text REtrieval Conference, TREC}}  (\bibinfo{year}{2021}).

\bibitem{pradeep2023pygaggle}
\bibinfo{author}{Pradeep, R.}, \bibinfo{author}{Chen, H.}, \bibinfo{author}{Gu, L.}, \bibinfo{author}{Tamber, M.~S.} \& \bibinfo{author}{Lin, J.}
\newblock \bibinfo{title}{Pygaggle: A gaggle of resources for open-domain question answering}.
\newblock \emph{\bibinfo{journal}{Advances in Information Retrieval: 45th European Conference on Information Retrieval}} \textbf{\bibinfo{volume}{13982}}, \bibinfo{pages}{148--162} (\bibinfo{year}{2023}).

\bibitem{nogueira2019passage}
\bibinfo{author}{Nogueira, R.} \& \bibinfo{author}{Cho, K.}
\newblock \bibinfo{title}{Passage re-ranking with {BERT}}.
\newblock \emph{\bibinfo{journal}{arXiv preprint arXiv:1901.04085}}  (\bibinfo{year}{2019}).

\bibitem{rosa2022no}
\bibinfo{author}{Rosa, G.~M.} \emph{et~al.}
\newblock \bibinfo{title}{No parameter left behind: How distillation and model size affect zero-shot retrieval}.
\newblock \emph{\bibinfo{journal}{arXiv preprint arXiv:2206.02873}}  (\bibinfo{year}{2022}).

\bibitem{dijkstra2022reading}
\bibinfo{author}{Dijkstra, R.}, \bibinfo{author}{Gen{\c{c}}, Z.}, \bibinfo{author}{Kayal, S.}, \bibinfo{author}{Kamps, J.} \emph{et~al.}
\newblock \bibinfo{title}{Reading comprehension quiz generation using generative pre-trained transformers}.
\newblock \emph{\bibinfo{journal}{CEUR WS Proceedings}} \textbf{\bibinfo{volume}{3192}}, \bibinfo{pages}{4--17} (\bibinfo{year}{2022}).

\bibitem{chuklin2015click}
\bibinfo{author}{Chuklin, A.}, \bibinfo{author}{Markov, I.} \& \bibinfo{author}{de~Rijke, M.}
\newblock \emph{\bibinfo{title}{Click Models for Web Search}}  (\bibinfo{year}{2015}).

\bibitem{ouyang2022training}
\bibinfo{author}{Ouyang, L.} \emph{et~al.}
\newblock \bibinfo{title}{Training language models to follow instructions with human feedback}.
\newblock \emph{\bibinfo{journal}{Advances in Neural Information Processing Systems}} \textbf{\bibinfo{volume}{35}}, \bibinfo{pages}{27730--27744} (\bibinfo{year}{2022}).

\bibitem{gpt4}
\bibinfo{author}{OpenAI}.
\newblock \bibinfo{title}{Gpt-4 technical report}.
\newblock \emph{\bibinfo{journal}{arXiv:submit/4812508}}  (\bibinfo{year}{2023}).

\bibitem{christophe2024med42}
\bibinfo{author}{Christophe, C.}, \bibinfo{author}{Kanithi, P.~K.}, \bibinfo{author}{Raha, T.}, \bibinfo{author}{Khan, S.} \& \bibinfo{author}{Pimentel, M.~A.}
\newblock \bibinfo{title}{Med42-v2: A suite of clinical llms}.
\newblock \emph{\bibinfo{journal}{arXiv preprint arXiv:2408.06142}}  (\bibinfo{year}{2024}).

\bibitem{cot2022}
\bibinfo{author}{Wei, J.} \emph{et~al.}
\newblock \bibinfo{title}{Chain-of-thought prompting elicits reasoning in large language models}.
\newblock \emph{\bibinfo{journal}{Advances in Neural Information Processing Systems}} \textbf{\bibinfo{volume}{35}}, \bibinfo{pages}{24824--24837} (\bibinfo{year}{2022}).

\bibitem{asai2023retrieval}
\bibinfo{author}{Asai, A.}, \bibinfo{author}{Min, S.}, \bibinfo{author}{Zhong, Z.} \& \bibinfo{author}{Chen, D.}
\newblock \bibinfo{title}{Retrieval-based language models and applications}.
\newblock \emph{\bibinfo{journal}{Proceedings of the 61st Annual Meeting of the Association for Computational Linguistics}} \bibinfo{pages}{41--46} (\bibinfo{year}{2023}).

\bibitem{guu2020retrieval}
\bibinfo{author}{Guu, K.}, \bibinfo{author}{Lee, K.}, \bibinfo{author}{Tung, Z.}, \bibinfo{author}{Pasupat, P.} \& \bibinfo{author}{Chang, M.}
\newblock \bibinfo{title}{Retrieval augmented language model pre-training}.
\newblock \emph{\bibinfo{journal}{International Conference on Machine Learning}} \bibinfo{pages}{3929--3938} (\bibinfo{year}{2020}).

\end{thebibliography}

\section*{Figure Legends}

Figure \ref{fig:cum-accuracy-all}: {\textbf{Precision@n obtained from the different search engines.} Each data point on the graphs represents the precision at the n-th position of the ranking (Precision@n) averaged over all queries in each collection. For example, the fourth point in each line (position=4) represents the average ratio of correct responses obtained after inspecting the top 4 results (P@4). 
Each line represents one search
engine (Bing is the blue line; Duckduckgo, the green line; Google, the red line; and Yahoo, the light blue line).
Panel a) plots results for the TREC HM 2020 collection. 
Panel b) plots results for the TREC HM 2021 collection. 
Panel c) plots results for the TREC HM 2022 collection.

\noindent Figure \ref{fig:ans}: {\textbf{Search Engine Answering Scores.} Average number of health questions (out of 50) that were answered at the top 20. Each engine is represented by an increasingly darker blue bar in the following order: Google, Bing, Yahoo, and Duckduckgo.

\noindent Figure \ref{fig:user-heuristic}: \textbf{Lazy and Diligent User Models}. Percentages of correct, incorrect, and no decision for the different datasets, search engines, and user models. Each panel is divided into two parts: the left side for lazy user behavior and the right side for diligent user behavior. Each bar presents the percentage of correct (green stacked bar), incorrect (red stacked bar), and no decision (grey stacked bar) for the corresponding user behavior.  
Panel a) plots results for the TREC HM 2020 collection. 
Panel b) plots results for the TREC HM 2021 collection. 
Panel c) plots results for the TREC HM 2022 collection.

\noindent Figure \ref{fig:zeroshot}: \textbf{Zero-shot experiments.} For each LLM, the bars represent the accuracy for the three prompting strategies: ``no context'' (light blue bar), ``non-expert'' (blue bar), and ``expert'' (navy blue bar).
Panel a) plots results for the TREC HM 2020 collection. 
Panel b) plots results for the TREC HM 2021 collection. 
Panel c) plots results for the TREC HM 2022 collection.

\noindent Figure \ref{fig:errors}: \textbf{Most common errors made by the LLMs}. Percentage of different types of errors over the total number of errors made by the ``no context'' (navy blue bar) and ``expert'' prompting (blue bar) strategies.

\noindent Figure \ref{fig:rag-no-context}: \textbf{Retrieval-augmented LLMs, accuracy for ``no context'' prompt}. Each ``Top n'' bar depicts the performance obtained by feeding the n-th result from Google until the top 5 (top 1 results are represented as a light blue bar, top 2 as a navy blue bar, top 3 as a light green bar, top 4 as a green bar and top 5 as a salmon bar). The baseline (no retrieval augmentation) is represented by a red dashed line. 
Panel a) plots results for the TREC HM 2020 collection. 
Panel b) plots results for the TREC HM 2021 collection. 
Panel c) plots results for the TREC HM 2022 collection.

\noindent Figure \ref{fig:rag-expert}: \textbf{Retrieval-augmented LLMs, accuracy for ``expert'' prompt}. Each ``Top n'' bar depicts the performance obtained by feeding the n-th result from Google until the top 5 (top 1 results are represented as a light blue bar, top 2 as a navy blue one, top 3 as a light green one,  top 4 as a green one and top 5 as a salmon one). The baseline (no retrieval augmentation) is represented by a red dashed line. 
Panel a) plots results for the TREC HM 2020 collection. 
Panel b) plots results for the TREC HM 2021 collection. 
Panel c) plots results for the TREC HM 2022 collection.

\noindent Figure \ref{fig:rag-influence}: \textbf{RAG experiments with 3 passages injected}. Accuracy of the LLMs with varying numbers of correct passages (0/3, 1/3, 2/3, or 3/3). On each panel, GPT-4 is represented by a blue line, LLama3 by an orange line, MedLlama3 by a green line, ChatGPT by a red line, and text-davinci-002 by a purple line.
Panel a) plots results for the TREC HM 2020 collection. 
Panel b) plots results for the TREC HM 2021 collection. 
Panel c) plots results for the TREC HM 2022 collection.

\noindent Figure \ref{fig:topic}: \textbf{Example of a Health Question}. Topic from the TREC Health Misinformation Track. The data fields used in the experiments are presented in red boxes.
   
\noindent Figure \ref{fig:se-architecture}: \textbf{Extracting responses for health questions from the search engine results}. The passages shown are real passages obtained from the top search results for the question ``Will wearing ankle braces help heal tendonitis?''. The process consists of: first, retrieving top 20 results for the query (provided by each search engine); second, selecting the most relevant passages from the retrieved pages (query-biased passages, using MonoT5); third, asking GPT-3 to read each passage and determine if it answers 'yes', 'no', or does not provide an answer to the health question.

\noindent Figure \ref{fig:user-models}: \textbf{Example of the user behavior models.} The lazy user makes a wrong decision because he stops right after seeing the first response (second document). In contrast, the diligent user makes the right decision after reading three responses (four documents). Note that an entry can give no response to the question and that the provided responses can also contain incorrect information.

\section*{Author contributions}

All authors contributed equally to this work as co-senior authors. Conceptualization: M.F, J.C., D.L.; experimentation: M.F; visualisation: M.F, J.C., D.L.; writing - original draft preparation: M.F, J.C., D.L.; writing - review \& editing:  M.F, J.C., D.L.

\section*{Competing interestes}

We declare that we have no competing interests as defined by Springer or other interests that might be perceived to influence the results and/or discussion reported in this paper.

\end{document}